# 3D-PRINTABLE HEIGHT MODELS FOR DC CIRCUITS

Oliver Bodensiek, Dörte Sonntag, Isabelle Glawe, Rainer Müller

Institute for Educational Research in Natural Sciences
Technische Universität Braunschweig, Germany

*We present an approach to learn about electricity and dc electric circuits by using real representational models based on the gravitational analogy. The additively manufactured models are deployed in a small-scale teaching intervention on electric potential and voltage in simple electric circuits. From the preliminary results considerable learning gains can be inferred.*

## *INTRODUCTION*

Electric circuits constitute a difficult subject area both in school and even on an undergraduate university level – many students fail to develop a coherent understanding of electric circuits. Some of the most common and persistent misconceptions are *local thinking* and *confusion of current and voltage* and their causal interrelation [1-2]. Local thinking, for instance, implies believing that a current splits evenly at any junction irrespective of the resistance of corresponding branches. *Sequential reasoning* [3] is another, yet related, misconception implying circuit elements would only have an effect on other devices if they were *behind* them in terms of the direction of current flow.

On a macroscopic level these misconceptions can partly be related to an inaccurate or even missing understanding of *global* concepts such as electric potential [4] or electric field and their relation to local processes. Indeed, students often use solely the concept of current [1]. On the other hand, an understanding of the microscopic domain of electrons and the ability to relate it to macroscopic domain of potential difference, current and energy seems to be rather important for a consistent picture of electric [5-6]. In the case of transients, an instruction based on the macro-micro relationship has been observed to lead to a superior understanding [7].

A need for visual representations of concepts such as energy, voltage and current on one side, and of microscopic processes on the other side arises naturally from these findings. In the best case, such visualisations are consistent across secondary education, allow for a gradual progression in levels of conceptual understanding, and do not lead to interferences in learning when trying to link these different conceptual levels. Such visualizations or models should thus take into account both macroscopic concepts and microscopic processes. In order to reduce the abstractness of concepts, real representational models appear as a desirable supplement to any course on electric circuits.

## THE GRAVITATIONAL ANALOGY

A possible approach to consider all of the aforementioned aspects is based on the gravitational analogy between mechanical and electrical potential energy as outlined in the following. The work $W_{12}$ to move a charge $q$ from point 1 to point 2 against a homogenous electric field $\vec{E}$ is

$$W_{12} = \int_1^2 \vec{F} \cdot \mathrm{d}\vec{s} = q \int_1^2 \vec{E} \cdot \mathrm{d}\vec{s} = E_{pot,2} - E_{pot,1} \tag{1}$$

where $\vec{F} = q\,\vec{E}$ denotes the force and the last equation is only valid as long as there is no time-varying magnetic field present, i.e. $\vec{\nabla} \times \vec{E} = 0$. In this case, an electric potential $\phi_{el}$ exists such that $\vec{E} = -\vec{\nabla}\phi_{el}$ and the integrals in (1) are path independent and can be equated to the difference of the potential energy of the charge *in* the field. Dividing by $q$ delivers an energy-related measure independent of charge quantifying the capability of the electric field to do work on a charged object between two points in space and thereby to change its potential energy *in* the field. Moreover, it provides a definition of the voltage $V_{12}$:

$$\frac{W_{12}}{q} = \frac{E_{pot,2} - E_{pot,1}}{q} = \phi_{el,2} - \phi_{el,1} \equiv V_{12} \tag{2}$$

In a standard introductory class the converted version $V_{12} = W_{12}/q$ of the above equation often serves as *definition* of voltage. Thereby, voltage is possibly understood by students such that it defines the amount of energy that each charge *carries* with it ("work per charge" in the sense of a rucksack model). Physically more correct, energy transport is done by the electromagnetic *field* in direction of the Poynting vector. The charges or specifically electrons are rather drifting *passively* in a circuit, guided by the electric field. Using the object-independent measure of potential difference can be used to define the notion of voltage in order to support this picture of passive electrons.

The above equations are formally similar to the purely mechanical case of a mass $m$ in a force field. Even though year 7 or year 8 students are usually not even concerned with the *homogeneous* gravitational *field* $\vec{g}$[1], they do have useful preconceptions to build upon, e.g. they usually have some intuitive knowledge of the connection between potential energy and height. In addition, the corresponding potential energy $E_{pot} = m \cdot g \cdot h$ with $g = |\vec{g}|$ is part of almost all basic mechanics courses in secondary school and a gravitational potential $\phi_{grav} = g \cdot h$ can thus be easily introduced as object-independent measure. In order to find a simple formal analogy to this gravitational potential, the case of a homogeneous and straight conductor needs to be considered. The equation relating electric field and potential difference then reduces to

$$\phi_{el,2} - \phi_{el,1} = |\vec{E}|\, s_{12} \tag{3}$$

where $s_{12}$ denotes the distance between the two points under consideration. In order to define an unique potential function, we set the point 1 the zero-point of the potential and thus, omitting the indices, $\phi_{el} = |\vec{E}|\, s$. The gravitational analogy is summarised in Figure 1. Using this analogy for a two-dimensional dc circuit the electric potential can be encoded in the $z$-component or third spatial dimension, as shown in Figure 2. Voltage drops can then be simply read off as height differences.

---

[1] We neglect any spatial dependence here.

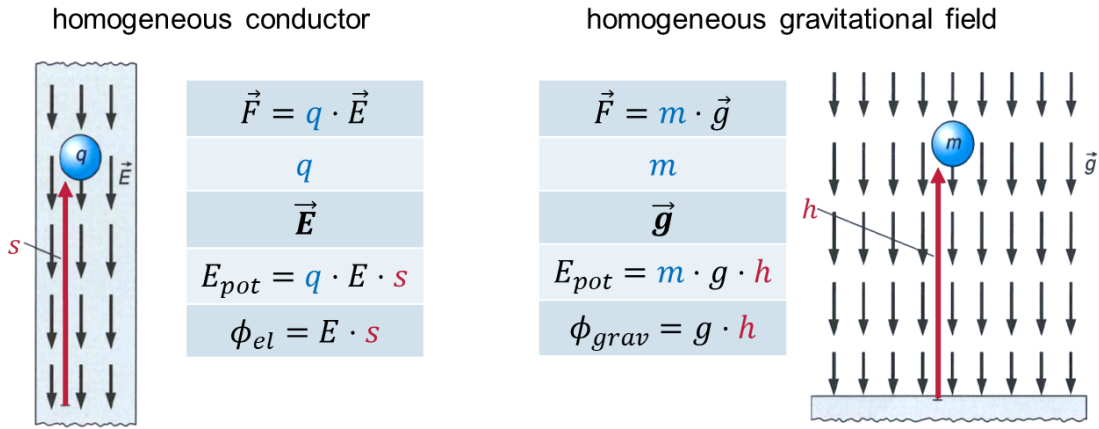

**Figure 1. The gravitational analogy between an electrical charge $q$ in a homogeneous electric field (left) and a mass $m$ in a homogeneous gravitational field (right), respectively.**

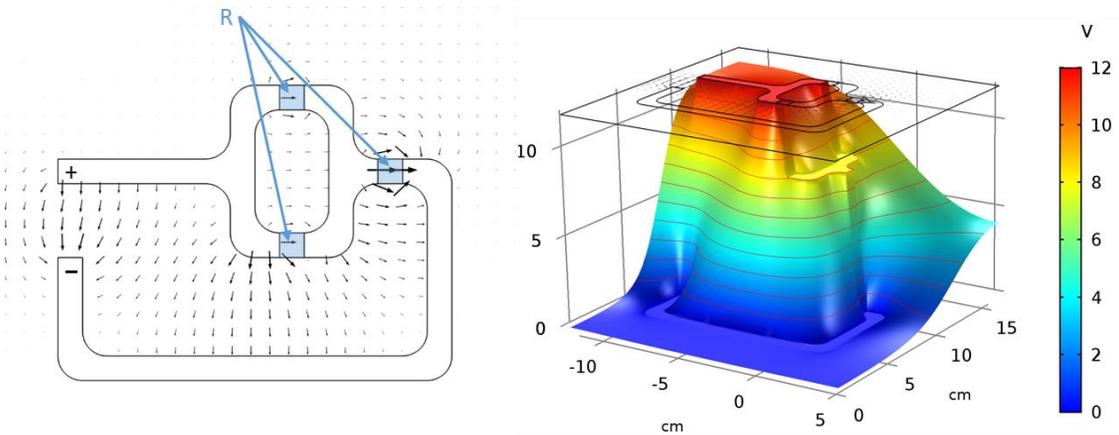

**Figure 2. Left: Mixed electric circuit (parallel and serial equal resistors, $V_{source} = 12\,V$). Black arrows indicate the electric field. Right: Corresponding electric potential encoded by height and colour. The circuit is plotted in black on top.**

## 3D-PRINTABLE HEIGHT MODELS

Although the global character of the electric potential field becomes evident from a visualisation as in Figure 2 where the potential is shown both in and around the conductor, a representation or model to be used in an introductory class should focus on the potential in a region confined to the conductor only. Such an early representational model for electric potential in dc circuits has been developed and investigated in [8]. In that model the terminals of each circuit element are represented by cylindrical elements on pillars, the height and colour of which correspond to the electric potential of the respective terminal. Voltage drops in circuit elements are visualised by the height difference of the two terminals whereas the terminals of different circuit elements on the same electric potential are horizontally interconnected by accordingly coloured cables.

We propose an improved height model, which we believe to be more intuitively accessible as it allows, for instance, for a continuous and direct visualisation of the gravitational analogy. In our approach, it is also possible to add a simple electron model by, e.g., metallic beads running along the pathway on the top of the model. However, a physical problem occurs here: Due to the negative charge $q = -e$, electrons would move upwards in the model from the negative to the positive terminal. A possible solution within the gravitational analogy is to redefine the electron charge as positive.

We utilized new additive manufacturing methods, which have become broadly accessible with the rise of commercially available 3D printers. For the case of height models for electric circuits, 3D printing offers new possibilities in design and construction and allows for a simple dissemination: For instance, a teacher in a school equipped with a 3D printer can simply download and print these models for teaching. Even a customization of parameters is possible to better fit to the experimental circuits used in each specific case. The printed representational model for a mixed circuit is shown in Figure 3 (right). We used a two-coloured printing to better visualise potential values of the physical circuit: The colour is alternating each centimetre in height corresponding to one volt.

In addition, we created dynamic animations of virtual models with the computer algebra system (CAS) Mathematica[2]. As additional teaching material, these animations can foster an even more intuitive access to the models as it is possible to completely fill the conductor pathway with beads (i.e. charges) in the virtual model and show them moving around, thereby representing a stationary current. A freeze frame of such an animation is shown in Figure 3 (left). Moreover, virtual applications in which parameter changes and the immediate visualisation of changes in the resulting height models and potential distributions, respectively, are possible.

In order to avoid the picture of solitarily moving electrons it is desirable to fill the pathway completely with beads also in the real models such that a collective and passive movement of electrons can be represented. Whereas it is easy to do so in the virtual case, it rather constitutes a challenge in the case of real models and is part of current development. In the real models shown here, only a few metallic beads can be moved along the channel on top of the models (c.f. Figure 3).

---

[2] The Mathematica code can be provided upon request.

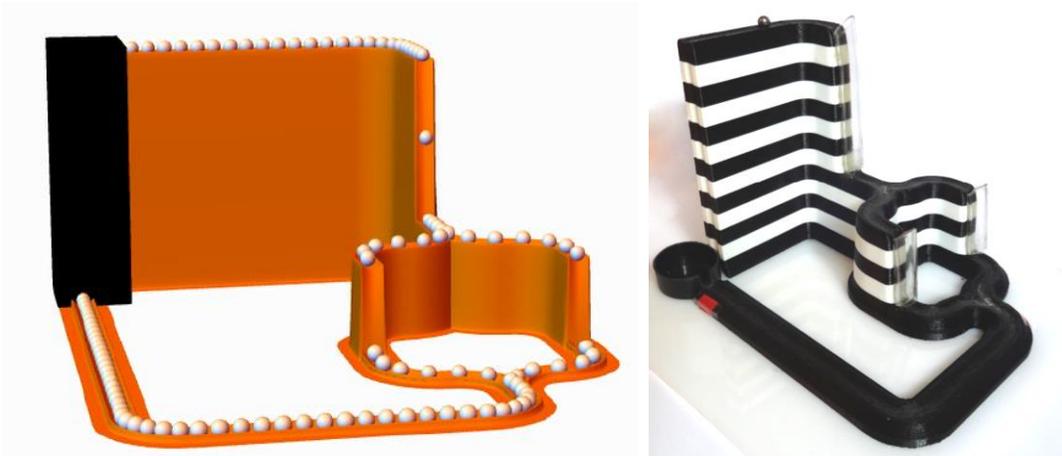

**Figure 3. Left: Freeze frame of the animation created in a CAS. Right: Printed height model.**

*EXPLORATORY TEACHING INTERVENTION*

**Intervention Design**
Using 3D-printed models for a series, a parallel and a mixed circuit, we have conducted an exploratory case study (N=59) with two year 8 courses in a pre-post design. We have, however, not included any control group or follow-up test at this stage. The students had been given lessons on electricity and electric circuits in a traditional curriculum before. Our preliminary results presented here have thus to be considered in the light of a possible effect of repetition of the subject matter. On the other hand, the intervention lasted only 90 minutes including 30 minutes for the post-test.

In the first 15 minutes, students have been introduced to the concept of potential (energy), the gravitational analogy and voltage as potential difference. In the following 35 minutes students worked in groups, each of which with one of the printed models (series, parallel, mixed). They obtained two worksheets: the first on potential and the second on voltage. They have been asked to sketch the electric circuit diagram, which corresponds to the height model, to colour parts of the circuit according to the electric potential and to compare it once again with the 3D model. Afterwards the student groups interchanged the height models in order to work with a different circuit topology. They have been asked again to sketch the corresponding circuit diagram and to colour the different equipotential parts. In the 3D model students ought to determine the voltages between several marked points. Afterwards they have compared the determined voltages to what they have measured in the corresponding real electric circuit. In the end students completed the post-test.

**Preliminary Results**
The primary objective of the preliminary teaching intervention was to test if a good understanding of the concept of potential and voltage can be achieved using the 3D-printed models presented above. Regarding electric potential, 36% of the students gave a correct definition after the teaching intervention, about 26% gave an incorrect or no definition and others gave only partly correct definitions. While 24% defined voltage either as "energy per charge" or formally by $U = W/q$ in the pre-test, this amount reduced to 10% in the post-test. Almost half of the students (44%) defined voltage as potential difference after the intervention. Many other students, who defined voltage differently, have however used the definition as potential difference subsequently in the post-test.

In one of the test items, students were asked to determine the voltage between different points in a simple circuit with one resistive element (light bulb) only, compare Figure 4.

**Figure 4. Test item on a simple electric circuit. Students are asked to determine the voltage between points A and B, B and C, C and D, and to reason their answers (analogous translation of the students' reasoning: "Since A and B, respectively C and D are lying on the same potential, the difference is zero.").**

In the pre-test 75% believed the source voltage of 4.5 V to be the voltage between all three pairs of points in the circuit. After the teaching intervention this answer was given by only 5% of the students. All three correct values were given by only 5% before and by 37% after the intervention. Even better, 73% of the students coloured the circuit correctly according to potential values.

A similar learning gain can be observed for a series circuit with two equal resistive elements, see Figure 5, where 29% identified a voltage drop of 3 V at each light bulb before the intervention and 63% afterwards. The correct determination of all four voltages (A-B, B-C, C-D, D-E) was only achieved by 3% before, but by 29% after the intervention. The correct colouring with respect to electric potential was done by 66%. All these findings are summarised in Table 1.

**Figure 5. Test item on a simple series circuit. Students are asked to determine the voltage drops at (a) and (b) between points A and B, B and C, C and D, and to reason their answers.**

**Table 1. Selected preliminary results of the teaching intervention as explained in the text.**

| test item | pre-test | post-test |
|---|---|---|
| definition of voltage | 24% correct | 70% correct |
| voltages at A-B, B-C, C-D (compare Figure 4) | 75% wrong | 5% wrong |
| determination of all three voltages (compare Figure 4) | 5% correct | 37% correct |
| voltages at resistances a, b (compare Figure 5) | 29 % correct | 63% correct |
| determination of all four voltages (compare Figure 5) | 3 % correct | 29% correct |

Contrary to these promising results, students had difficulties to solve similar problems for parallel or mixed circuits correctly. In these cases we do not observe a significant improvement following the teaching intervention. We believe that this is mostly due to the limited teaching time available in the preliminary study. In addition, students still had problems to differentiate between electric potential and voltage. This might be due to the almost simultaneous introduction of both concepts due to the limited amount of time available for the intervention. In a complete teaching unit based on the ideas outlined above, a distinct and even temporal separation between teaching the two concepts of electric potential and voltage should be made.

During the teaching intervention most of the students were particularly interested in working with the 3D-printed height models. They had no problems concerning the interpretation of the models as seen from the answers given on worksheets. The post-test with 81% of the students giving a reasonable explanation of the model seems to confirm this, too. Moreover, 92% of the students accept the model as a helpful tool and suggest its usage in physics education. About 75% believe that their understanding of voltage concept has increased. Matching the results from pre- and post-test, we do indeed observe considerable improvements in the understanding of the voltage concept.

## *DISCUSSION AND OUTLOOK*

We have presented an approach for learning and teaching electricity and dc electric circuits using representational, 3D-printable models based on the gravitational analogy, i.e. on potential energy. Due to the students' (pre-) conceptions of potential energy in the gravitational field, the models facilitate the transfer of the energy concept to electric circuits. The approach seems to be promising with respect to learning gains, especially relating to the concept of voltage. As the electron concept can be visualised in the models as well, a clear distinction between current and voltage can be made. As a further result not mentioned above, we observe almost no confusion of current and voltage in the post-test. Furthermore, it is, to a certain amount, possible to emphasise macro-micro relations using the models in courses on higher educational levels. Due to the formal analogies in theory, it is possible to use the gravitational analogy and even the corresponding models consistently throughout physics or even electrical engineering in secondary education. For instance, in higher courses the models can be extended such that the potential distribution and electric field in the entire space around the circuit is discussed.

A complete teaching unit, which includes the concept of current and different circuit topologies as well, based on the presented approach is currently under development. Learning effects will be investigated by an accompanying long-term study. We believe that the positive effects mentioned above can be confirmed in a statistically significant way. Additionally we are developing an appropriate model of the electrical source.

In conclusion, we found promising results indicating that a course based on the gravitational analogy in combination with representational models can considerably improve the understanding of electric circuits. Nevertheless, further research in this area is required.

*ACKNOWLEDGEMENTS*

The authors acknowledge funding by the German Federal Ministry of Education and Research. This project is part of the "Qualitätsoffensive Lehrerbildung", a joint initiative of the Federal Government and the *Länder* which aims to improve the quality of teacher training. The programme is funded by the Federal Ministry of Education and Research. The authors are responsible for the content of this publication.